# THE DEVELOPMENT OF A CHEMICAL KINETIC MECHANISM FOR COMBUSTION IN SUPERCRITICAL CARBON DIOXIDE

James Harman-Thomas, Kevin J. Hughes*, Mohamed Pourkashanian

Energy 2050, Department of Mechanical Engineering, The University of Sheffield, Sheffield, United Kingdom, S3 7RD.

**Corresponding author:* Kevin J. Hughes
*Email:* k.j.hughes@sheffield.ac.uk

## ABSTRACT

Direct fired supercritical $CO_2$ ($sCO_2$) power cycles allow for the combustion of gaseous fuels under oxyfuel conditions with inherent carbon capture. As the $CO_2$ is captured intrinsically, the efficiency penalty of capture on the overall plant is small, meaning that power plants achieve a similar efficiency to traditional fossil fuel power plants without carbon capture and storage. However, at high pressures and in large dilutions of $CO_2$, combustion mechanisms are poorly understood. Therefore, in this paper sensitivity and quantitative analysis of four established chemical kinetic mechanisms have been employed to determine the most important reactions and the best performing mechanisms over a range of different conditions. $CH_3O_2$ chemistry was identified as a pivotal mechanism component for modelling methane combustion above 200 atm. The University of Sheffield (UoS) $sCO_2$ mechanism created in the present work better models the ignition delay time (IDT) of high-pressure combustion in a large dilution of $CO_2$. Quantitative analysis showed that the UoS $sCO_2$ mechanism was the best fit to the greatest number of IDT datasets and had the lowest average absolute error value, thus indicating a superior performance compared to the four existing chemical kinetic mechanisms, well-validated for lower pressure conditions.

## ABBREVIATIONS

ASU – Air Separation Unit
*E* – Average Absolute Error
IDT – Ignition Delay Time
MWe – Megawatt Equivalent
$sCO_2$ – Supercritical $CO_2$
STEP – Supercritical Transformational Electric Power
UoS – University of Sheffield

## 1. INTRODUCTION

The global atmospheric carbon dioxide ($CO_2$) concentration has rapidly increased since the industrial revolution due to anthropogenic emissions and the atmospheric $CO_2$ concentration now exceeds 410 parts per million [1]. The power industry is a significant contributor to $CO_2$ emissions, due to a continued reliance on fossil fuels to power homes and businesses. Despite renewable energy becoming increasingly competitive, humanity's consumption of fossil fuels is not declining quickly enough, and thus new technologies to produce clean, emissions-free energy from fossil fuels are urgently required [2].

One emerging technology is direct-fired supercritical $CO_2$ ($sCO_2$) combustion which uses natural gas, or synthesis gas (syngas) produced from coal gasification, to produce electricity which inherently captures 100% of its emissions [3]. In a comparative review of emerging carbon capture and storage technologies, the Allam-Fetvedt cycle was the only coal combustion technology investigated which has the potential to reduce the production cost of electricity [4]. The Allam-Fetvedt cycle is a basic thermodynamic cycle that burns the gaseous fuel and high-purity oxygen in a $CO_2$ dilution of up to 96% [3]. Following combustion, the water produced is removed to leave high-purity, pipeline-ready $CO_2$, most of which is recycled into the combustion chamber with the small amount produced from combustion being removed. The high pressures of the Allam-Fetvedt cycle at the turbine inlet, coupled with the greater power density of the $sCO_2$ working fluid compared to traditional steam or gas working fluids leads to Allam-Fetvedt cycle power plants having a smaller footprint [3]. The Allam-Fetvedt cycle achieves high efficiency by alleviating the energy penalty of the air separation unit (ASU) by utilising the waste heat from compression in the ASU in the main power cycle [5]. A more detailed description and schematic of the natural gas-fired Allam-Fetvedt cycle power plant can be found in Allam et al. [6]. An alternative direct-fired $sCO_2$ power cycle is the Supercritical Transformational Electric Power (STEP) cycle which is currently being actualised as a 10 MWe pilot plant in San Antonio Texas, and this is set to be commissioned in 2022 [7].

A major challenge facing the advancement of direct-fired $sCO_2$ power cycles is a lack of experimental data at high pressure and a large dilution of $sCO_2$ [8]. Although current chemical kinetic mechanisms are well-validated at low-pressures and low-$CO_2$ dilutions, at direct-fired $sCO_2$ cycle conditions, there is little experimental data available to validate the kinetic mechanisms. Chemical kinetic mechanisms are required to be able to accurately model combustion and maximise the efficiency of the plant design. The present study evaluates how existing chemical kinetic mechanisms predict experimental data at large $CO_2$ dilutions for different gaseous fuels and identifies which reactions have the greatest influence on the ignition delay time (IDT). A series of sensitivity analyses were then used to assist in creating a new mechanism, the UoS $sCO_2$ that accurately models low and high-pressure oxyfuel combustion of hydrogen, syngas, and methane. Having a mechanism that accurately models the high-pressure combustion of the Allam-Fetvedt cycle is essential to improving the efficiency of the combustion chamber.

## 2. DATASET AND MECHANISM SELECTION

### 2.1. OXY-FUEL COMBUSTION

The present study focused on methane, hydrogen, and syngas as these are the most important fuels in the Allam-Fetvedt cycle. The datasets used in the present study were selected due to their high-pressure and large dilution of $CO_2$. All IDTs were recorded using the shock tube experimental technique which creates virtually instantaneous, high-pressure, adiabatic conditions. The reaction progress is monitored using various diagnostic techniques. Shock tubes do have some non-instantaneous pressure rise post-shock [9], which influences the IDT, but this has been omitted from the present study due to the data not being readily available for all of the datasets. Having a large dilution of $CO_2$ has both physical and chemical effects on combustion. As well as having a different third body efficiency to other bath gases, $CO_2$ may have a catalytic effect on some of the reactions [10]. In total, this study investigated 52 different datasets for oxyfuel combustion of methane, hydrogen, and syngas in various $CO_2$ dilutions at different pressures and equivalence ratios. To the best of the author's knowledge, this is the first mechanism created to model supercritical $CO_2$ combustion which studies such a comprehensive number of datasets for more than one fuel.

### 2.2. METHANE COMBUSTION

All available datasets for the IDT of methane in any dilution of $CO_2$ were collated and investigated in the present study. The datasets studied are from Hargis and Petersen [11], Koroglu et al. [12], Pryor et al. [13, 14], Liu et al. [15], Shao et al. [9], Karimi et al. [16], and Barak et al. [17]. Table 1 shows the twenty-nine datasets and the conditions that each dataset was recorded at. The average pressure of the datasets varies from sub-atmospheric to 266 atm, at various $CO_2$ dilutions and equivalence ratios, giving a comprehensive overview of oxy-methane combustion.

**Table 1.** Methane datasets analysed in the present study.

| Dataset | Reference | Average Pressure /atm | Equivalence Ratio (Φ) | $CO_2$ Dilution (%) |
|---|---|---|---|---|
| M1 | [11] | 1.7 | 2.00 | 50.00 |
| M2 | [11] | 2.5 | 2.00 | 50.00 |
| M3 | [11] | 2.1 | 2.00 | 75.00 |
| M4 | [11] | 12.7 | 2.00 | 75.00 |
| M5 | [12] | 0.8 | 1.00 | 30.00 |
| M6 | [12] | 3.8 | 1.00 | 30.00 |
| M7 | [12] | 0.8 | 2.00 | 30.00 |
| M8 | [12] | 3.9 | 2.00 | 30.00 |
| M9 | [12] | 0.7 | 0.50 | 30.00 |
| M10 | [12] | 3.6 | 0.50 | 30.00 |
| M11 | [12] | 0.6 | 1.00 | 60.00 |
| M12 | [13] | 0.9 | 1.00 | 60.00 |
| M13 | [13] | 7.2 | 1.00 | 60.00 |
| M14 | [13] | 8.9 | 1.00 | 85.00 |
| M15 | [13] | 29.6 | 1.00 | 85.00 |
| M16 | [14] | 0.6 | 1.00 | 89.50 |
| M17 | [14] | 1.0 | 1.00 | 85.00 |
| M18 | [15] | 1.9 | 2.00 | 75.00 |
| M19 | [15] | 0.9 | 2.00 | 75.00 |
| M20 | [9] | 32.2 | 1.00 | 77.50 |
| M21 | [9] | 106.3 | 1.00 | 77.50 |
| M22 | [9] | 260.0 | 1.00 | 77.50 |
| M23 | [9] | 31.4 | 1.27 | 86.17 |
| M24 | [9] | 74.7 | 1.27 | 86.17 |
| M25 | [9] | 266.3 | 1.27 | 86.17 |
| M26 | [16] | 99.0 | 1.00 | 85.00 |
| M27 | [16] | 97.0 | 0.50 | 80.00 |
| M28 | [16] | 201.8 | 1.00 | 85.00 |
| M29 | [17] | 79.9 | 1.00 | 36.50 |

### 2.3. HYDROGEN COMBUSTION

The only published IDT data for hydrogen combustion in $CO_2$ is by Shao et al. [9]. The authors report an IDT value for hydrogen combustion diluted in 85% $CO_2$ at two different equivalence ratios over a large pressure range as shown in Table 2.

**Table 2.** Hydrogen datasets analysed in the present study.

| Dataset | Reference | Average Pressure /atm | Equivalence Ratio (Φ) | $CO_2$ Dilution (%) |
|---|---|---|---|---|
| H1 | [9] | 109.6 | 1.00 | 85.00 |
| H2 | [9] | 270.6 | 1.00 | 85.00 |
| H3 | [9] | 38.4 | 0.25 | 85.00 |

### 2.4 SYNTHESIS GAS COMBUSTION

Oxy-syngas combustion data published in Vasu et al. [18], Karimi et al. [19] and Barak et al. [17, 20, 21] was investigated during the present study as shown in Table 3. This collection of datasets covers a large range of $CO_2$ dilutions and equivalence ratios, at average pressures ranging from 1.2 to 208 atmospheres.

**Table 3.** Syngas datasets analysed in the present study.

| Dataset | Reference | Average Pressure /atm | Equivalence Ratio (Φ) | $CO_2$ Dilution (%) |
|---|---|---|---|---|
| S1 | [18] | 1.2 | 1.00 | 24.44 |
| S2 | [18] | 1.7 | 1.00 | 24.44 |
| S3 | [18] | 2.3 | 1.00 | 24.44 |
| S4 | [20] | 1.7 | 0.50 | 60.00 |
| S5 | [20] | 1.7 | 0.50 | 80.00 |
| S6 | [20] | 1.7 | 0.33 | 75.00 |
| S7 | [20] | 1.7 | 1.00 | 85.00 |
| S8 | [20] | 1.7 | 0.50 | 80.00 |
| S9 | [20] | 1.7 | 0.50 | 80.00 |
| S10 | [21] | 41.5 | 1.00 | 85.00 |
| S11 | [21] | 38.6 | 1.00 | 85.00 |
| S12 | [21] | 38.5 | 1.00 | 85.00 |
| S13 | [21] | 38.4 | 1.00 | 85.00 |
| S14 | [17] | 78.9 | 1.02 | 91.80 |
| S15 | [17] | 91.7 | 0.41 | 64.50 |
| S16 | [17] | 89.6 | 0.41 | 92.20 |
| S17 | [17] | 89.7 | 1.09 | 63.90 |
| S18 | [19] | 101.0 | 1.00 | 95.50 |
| S19 | [19] | 84.2 | 2.00 | 92.50 |
| S20 | [19] | 208.0 | 1.00 | 95.50 |

## 2.5. MECHANISM SELECTION

The four mechanisms used in this study were selected based on their suitability for modelling the combustion of lower hydrocarbons. Aramco 2.0 which was originally released in 2013 and has been subsequently updated contains the greatest amount of chemistry of the selected mechanisms with 493 species and 2716 reactions and was developed for the combustion of $C_1$-$C_4$ hydrocarbons and hydrogen [22-28]. The DTU mechanism released in 2019 contains 102 species and 894 reactions and was developed by the Technical University of Denmark for the high-pressure combustion of $H_2$ and $C_1$/$C_2$ hydrocarbons [29, 30]. The GRI 3.0 mechanism was originally released in 1999 and contains 53 species and 325 reactions and is validated against IDTs for methane and ethane below 100 atm [31]. USC II was released in 2007 and contains 111 species and 784 reactions and is applicable to the combustion of $H_2$/CO/$C_1$-$C_4$ compounds [32].

The mechanisms are quantitatively evaluated using an average absolute error ($E$,%) value using the following expression [33]:

**Eq. (1)** $$E(\%) = \frac{1}{N}\sum_{i=1}^{N}\left|\frac{X_{sim,i}-X_{exp,i}}{X_{exp,i}}\right| \times 100$$

Where $N$ represents the number of data points in the experimental set, $X_{sim,i}$ and $X_{exp,i}$ are the modelled and experimental results for the $i$th IDT data point respectively. A mechanism's $E$ value gives a quantitative indication of performance and the smaller the $E$ value, the smaller the difference between the experimental and measured data point and the better the mechanism performs.

It must be noted that due to the error in the data points, the quantitative analysis cannot be considered a definitive way to determine which mechanism is best performing when multiple mechanisms lie within the error bounds. However, due to the large number of datasets being investigated, it is a good indicator of the mechanism's performance.

## 3. MODELLING PROCEDURE

The modelling work undertaken throughout this study was performed using ANSYS Chemkin-Pro 2019 R3 [34]. The IDT data was modelled for different test gases using four existing chemical mechanisms. A closed homogeneous batch reactor with the 'constrain volume and solve energy equation' problem type which resembles the adiabatic conditions of the test-gas region of the shock tube was used to model the IDT for all the datasets. To determine the IDT, the mole fraction of OH was plotted against the reaction time at a given temperature. The IDT was defined as the difference between time-zero and the onset of ignition, the onset of ignition being defined as the time of the maximum gradient of the increasing OH concentration. Any errors shown on the IDT plots of the datasets were taken from the original journal publications and varied from 18 to 25%.

A series of sensitivity analyses were performed using OH species sensitivity at the IDT for a given temperature within a dataset. The temperatures were selected at points of large discrepancy between either the mechanism and the experimental data point or the different mechanisms being studied. The top ten most sensitive reactions were determined and plotted as a bar chart where increasing the reaction rate would reduce the IDT for a reaction with a positive sensitivity coefficient and increase the IDT for a reaction with a negative sensitivity coefficient. The key equations used with the Chemkin software and the sensitivity analysis can be found in the ANSYS Chemkin Theory Manual 17.0 (15151) [35].

The new University of Sheffield (UoS) $sCO_2$ mechanism was created by interpreting the sensitivity analysis where the mechanism performed well and comparing the rate coefficients of the four mechanisms studied. This allowed the selection of the best rate coefficient to better model every dataset, without affecting the conditions where the mechanism performed well.

### 3.1. ANALYSIS OF METHANE DATASETS

Table 4 compares the four mechanisms ability to model each of the 29 experimental methane IDT datasets using equation 1. The penultimate row shows the average, average absolute error value (average *E*) across all the datasets and the final row counts how many times each mechanism was the best fit to a dataset. It was found that the DTU has the lowest average *E* value of 23.89%, closely followed by Aramco 2.0 with 28.89%. Interestingly, GRI 3.0 and USC II which are the best fit to the greatest number of datasets, have the largest average *E* value. This implies the two mechanisms have a large average absolute error in a small number of conditions which significantly increase the average *E* value. This proves to be true as in M22, M25 and M28 where GRI 3.0 and USC II have a much larger average absolute error compared to Aramco 2.0 and DTU. These 3 datasets were the highest-pressure conditions studied, all recorded at an average pressure above 200 atm. This suggests that there is a flaw in the GRI 3.0 and USC II chemical kinetic mechanisms which causes a significant over prediction of IDTs above 200 atm.

**Table 4.** Quantitative analysis of the methane datasets.

| Dataset | Average Absolute Error (*E*, (%)) |
|---|---|

| No. | Reference | Average Pressure /atm | Aramco 2.0 | DTU | GRI 3.0 | USC II |
|---|---|---|---|---|---|---|
| **M1** | [11] | 1.7 | 44.4 | 19.3 | 13.6 | 14.4 |
| **M2** | [11] | 2.5 | 15.2 | 11.9 | 31.6 | 23.7 |
| **M3** | [11] | 2.1 | 29.6 | 12.1 | 16.7 | 9.2 |
| **M4** | [11] | 12.7 | 15.0 | 19.0 | 50.4 | 34.8 |
| **M5** | [12] | 0.8 | 15.1 | 9.5 | 3.7 | 14.4 |
| **M6** | [12] | 3.8 | 25.0 | 17.4 | 14.1 | 15.5 |
| **M7** | [12] | 0.8 | 25.8 | 25.4 | 11.8 | 24.6 |
| **M8** | [12] | 3.9 | 40.2 | 36.8 | 13.9 | 19.3 |
| **M9** | [12] | 0.7 | 13.5 | 6.2 | 7.2 | 8.5 |
| **M10** | [12] | 3.6 | 32.2 | 19.5 | 11.9 | 15.7 |
| **M11** | [12] | 0.6 | 11.2 | 10.3 | 5.9 | 33.8 |
| **M12** | [13] | 0.9 | 30.6 | 23.9 | 12.2 | 32.2 |
| **M13** | [13] | 7.2 | 30.2 | 23.9 | 17.6 | 16.3 |
| **M14** | [13] | 8.9 | 23.0 | 17.7 | 16.0 | 5.1 |
| **M15** | [13] | 29.6 | 35.5 | 31.4 | 34.7 | 12.3 |
| **M16** | [14] | 0.6 | 34.0 | 19.4 | 44.7 | 114.6 |
| **M17** | [14] | 1.0 | 18.3 | 14.3 | 9.9 | 13.1 |
| **M18** | [15] | 1.9 | 51.3 | 32.2 | 12.1 | 14.7 |
| **M19** | [15] | 0.9 | 71.1 | 56.1 | 27.6 | 38.6 |
| **M20** | [9] | 32.2 | 60.6 | 47.2 | 23.8 | 39.0 |
| **M21** | [9] | 106.3 | 14.1 | 16.2 | 43.1 | 21.7 |
| **M22** | [9] | 260.0 | 13.2 | 10.0 | 206.8 | 328.0 |
| **M23** | [9] | 31.4 | 94.5 | 98.8 | 31.9 | 59.9 |
| **M24** | [9] | 74.7 | 20.7 | 24.8 | 56.4 | 10.6 |
| **M25** | [9] | 266.3 | 11.2 | 30.8 | 31.9 | 131.9 |
| **M26** | [16] | 99.0 | 24.7 | 26.6 | 54.2 | 21.8 |
| **M27** | [16] | 97.0 | 14.3 | 13.1 | 56.7 | 7.9 |
| **M28** | [16] | 201.8 | 7.1 | 6.5 | 32.7 | 45.4 |
| **M29** | [17] | 79.9 | 16.9 | 12.6 | 54.4 | 9.6 |
| ***Average E (%)*** | | | ***28.9*** | ***23.9*** | ***32.7*** | ***39.2*** |
| ***No. Best Fit*** | | | ***3*** | ***5*** | ***13*** | ***8*** |

Figure 1 shows the sensitivity analysis of the M25 condition at 1100 K for Aramco 2.0 and USC II which has a significant discrepancy between the modelled IDTs. One of the initial observations was the presence of the $CH_3O_2$ and $CH_3O_2H$ species within the Aramco 2.0 sensitivity analysis which were absent in USC II. This is because USC II, as well as GRI 3.0, do not contain $CH_3O_2$ and $CH_3O_2H$, or any of their respective chemistry. In the paper in which the data was published, Shao et al. [9], modelled the data with Aramco 2.0 and FFCM-1 [36] and found a similar agreement between modelled IDTs at the lower pressure conditions and a large discrepancy above 200 atm. This is consistent with the current findings as the FFCM-1 mechanism does not contain $CH_3O_2$ and $CH_3O_2H$.

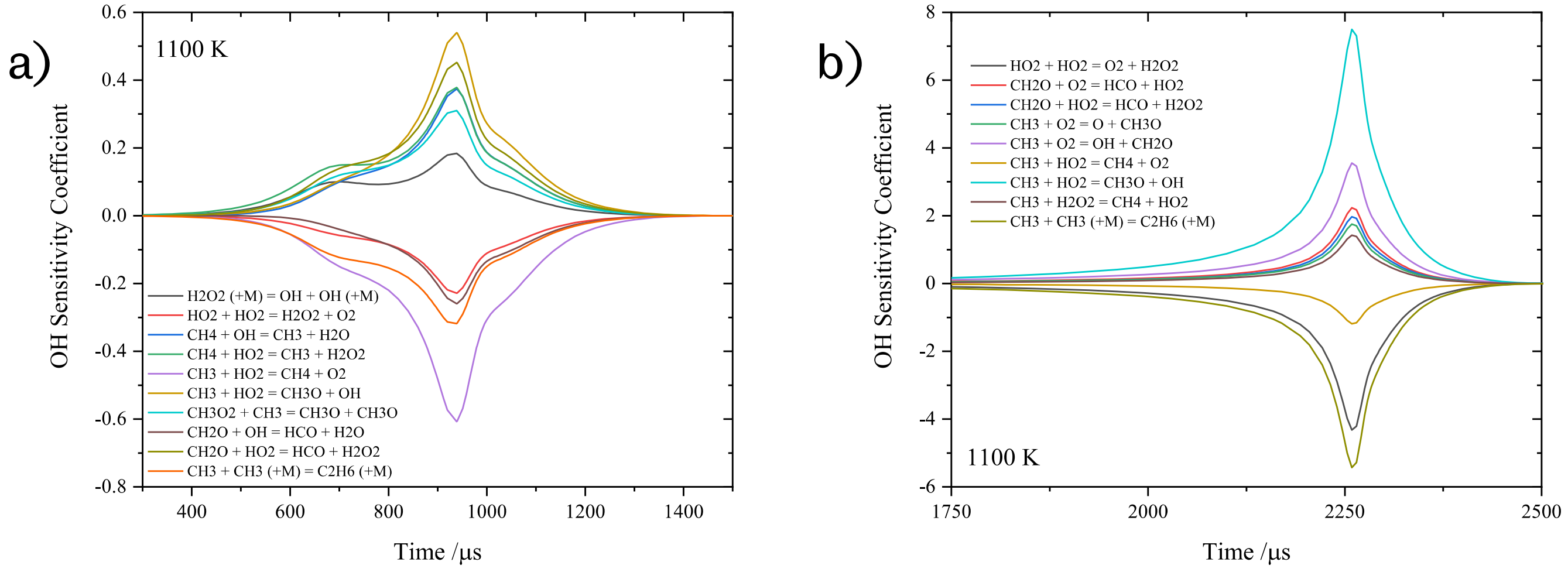

**Figure 1:** Plot of OH sensitivity coefficient as a function of time for dataset M25; a) Aramco 2.0 and b) USC II.

This observation is key to creating a new mechanism that can model combustion in $sCO_2$. The formation of $CH_3O_2$ via the recombination of $CH_3$ and $O_2$ as shown in R1 has most recently been studied by Fernandes et al. [37] using a high-temperature, high-pressure flow cell in an argon and nitrogen bath gas. Although the authors do not report an error, they are in good agreement with existing experimental data. Therefore, the rate coefficient can be considered well known at temperatures below 700 K. However, none of this research was performed with a $CO_2$ bath gas, so despite the good agreement between the different bath gases studied, the rate coefficient may be different in a large dilution of CO and at higher temperatures.

R1 $CH_3 + O_2 (+M) \rightleftharpoons CH_3O_2 (+M)$

There are two reactions of $CH_3O_2$ that appear in the sensitivity analysis of Aramco 2.0 for M22, M25 and M28: R2 and R3.

R2 $CH_4 + CH_3O_2 \rightleftharpoons CH_3 + CH_3O_2H$

R3 $CH_3O_2 + CH_3 \rightleftharpoons CH_3O + CH_3O$

The rate coefficient of R2 used in Aramco 2.0 is from an unknown and unpublished source and therefore it is difficult to evaluate. The only published value of the rate coefficient of R2 comes from a review of chemical kinetic data of methane and related compounds by Tsang and Hampson [38]. The authors note there is no direct measurement and thus base their value on R4, a reaction which they argue should have a similar rate coefficient. Due to it being an estimated rate coefficient, there is a large amount of uncertainty in the rate coefficient used for R2.

R4 $HO_2 + CH_4 \rightleftharpoons CH_3 + H_2O_2$

The rate coefficient used for R3 in Aramco 2.0 was theoretically calculated by Keiffer et al. [39] at 0.169 bar from 298-530 K in an oxygen bath gas. The reaction has been measured twice experimentally by Pilling and Smith [40] and Parkes [41] in argon and nitrogen and bath gases respectively, both at 298 K. Due to the importance of R3 in the high-pressure combustion of methane, it is important to revisit this reaction at conditions more relevant to direct-fired $sCO_2$ combustion.

Other reactions relevant to the $CH_3O_2$ chemistry used in Aramco 2.0 use rate coefficients from Lightfoot et al. [42] between 600K and 719 K at atmospheric pressure. Previous research into $CH_3O_2$ kinetics has focused on atmospheric and low-temperature combustion chemistry [43].

In addition to the importance of the $CH_3O_2$ chemistry, two other discrepancies between the sensitivity analysis of Aramco 2.0 and USC II for the M22 and M25 conditions were identified. Firstly, R5 appeared only for Aramco 2.0. The USC II mechanism uses an older rate coefficient from Baulch et al. [44] whereas Aramco 2.0 uses a more recent theoretical rate coefficient from Troe et al. [45] which was calculated with a $CO_2$ bath gas.

R5 $H_2O_2\ (+M) \rightleftharpoons OH + OH\ (+M)$

The second discrepancy noted was the Aramco 2.0 mechanism was considerably more sensitive to R6 than USC II. For this reaction, the USC II rate coefficient is taken from Reid et al. [46] which is reported to be private communication. The theoretically calculated Aramco 2.0 rate coefficient from Jasper et al. [47] is in agreement with Srinivasan et al. [48] which incorporated experimental data.

R6 $CH_3 + HO_2 \rightleftharpoons CH_4 + O_2$

The effect of altering the rate coefficients for R5 and R6 and adding $CH_3O_2$ chemistry into the USC II mechanism was investigated. These changes were made to the USC II mechanism sequentially as shown in figure 2 for conditions M22 and M25.

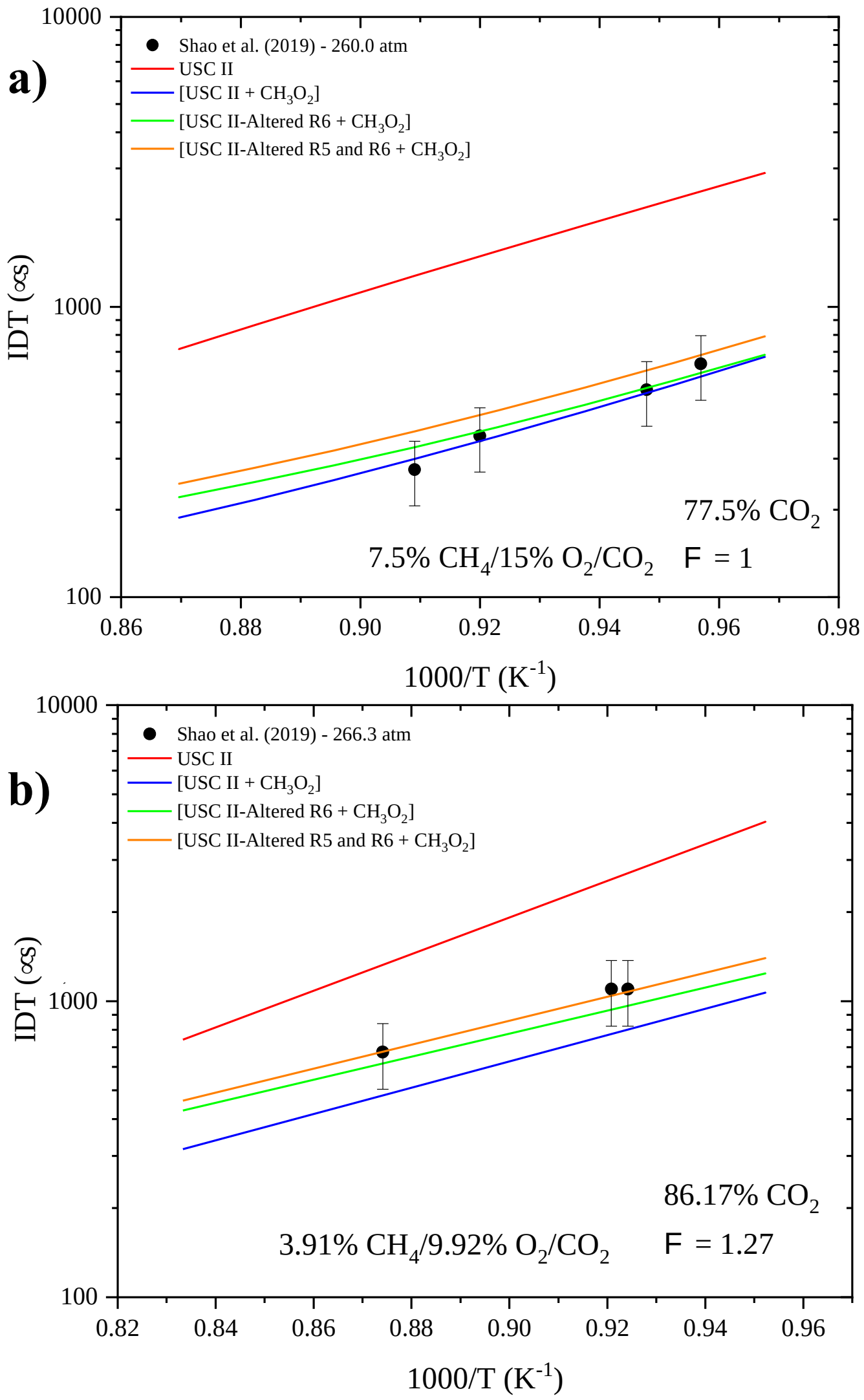

**Figure 2:** Sequential changes to USC II; a) M22 and b) M25; [USC II + $CH_3O_2$]: addition of $CH_3O_2$ chemistry from Aramco 2.0, [USC II-Altered R6 + $CH_3O_2$] change R6 to Aramco 2.0 rate coefficient, [USC II-Altered R5 and R6 + $CH_3O_2$]: change R5 to the Aramco 2.0 rate coefficient.

Figure 2 highlights the importance of the chemistry of $CH_3O_2$ in high-pressure combustion. The addition of the following reactions shown in Table 5 and the respective thermodynamic and transport data to the USC II mechanism immediately led to a significant improvement in the mechanism's ability to model the experimental data. This observation is significant in the pursuit of understanding the chemical kinetic mechanism of direct-fired $sCO_2$ combustion. However, despite the importance of the reactions shown in Table 5, these rate coefficients are often from sources concerned with atmospheric and low-temperature combustion and therefore are difficult to extrapolate to the combustion conditions of direct fired $sCO_2$ cycles. Furthermore, for some of the rate coefficients, even for these inappropriate conditions there are large uncertainty factors that could have a huge impact on the modelled IDTs. For example, Tsang and Hampson [38] report an uncertainty factor of 10 in the A factor for the rate coefficient of R2 due to it being an estimate based on another reaction with no experimental data. Therefore, to create an accurate comprehensive kinetic mechanism of high-pressure combustion, the rate coefficients of the key reactions in the chemistry of $CH_3O_2$ must be determined at larger pressures and temperatures.

**Table 5.** Reactions of $CH_3O_2$ added to USC II.

| Reaction | A (cm$^3$ mol s) | n | Ea (cal/mol) | Reference |
|---|---|---|---|---|
| **$CH_3O_2 + CH_3 \rightleftharpoons 2CH_3O$** | 5.08 x $10^{12}$ | 0.00 | -1411 | [39] |
| **$CH_4 + CH_3O_2 \rightleftharpoons CH_3 + CH_3O_2H$** | 9.60 x $10^{-01}$ | 3.77 | 17810 | Unknown |
| **$CH_2O + CH_3O_2 \rightleftharpoons HCO + CH_3O_2H$** | 1.99 x $10^{12}$ | 0.00 | 11660 | [38] |
| **$CH_3 + O_2 (+M) \rightleftharpoons CH_3O_2 (+M)$** | 7.81 x $10^{9}$ | 0.90 | 0 | [37] |
| **$CH_3O_2 + O \rightleftharpoons CH_3O + O_2$** | 3.60 x $10^{13}$ | 0.00 | 0 | [42] |
| **$CH_3O_2 + H \rightleftharpoons CH_3O + OH$** | 9.60 x $10^{13}$ | 0.00 | 0 | [42] |
| **$CH_3O_2 + OH \rightleftharpoons CH_3OH + O_2$** | 6.00 x $10^{13}$ | 0.00 | 0 | [42] |
| **$CH_3O_2 + HO_2 \rightleftharpoons CH_3O_2H + O_2$** | 2.47 x $10^{11}$ | 0.00 | -1570 | [42] |
| **$CH_3O_2 + H_2O_2 \rightleftharpoons CH_3O_2H + HO_2$** | 2.41 x $10^{12}$ | 0.00 | 9936 | [38] |
| **$2CH_3O_2 \rightleftharpoons CH_2O + CH_3OH + O_2$** | 3.11 x $10^{14}$ | -1.61 | -1051 | [42] |
| **$CH_3O_2 + CH_3O_2 \rightleftharpoons O_2 + CH_3O + CH_3O$** | 1.40 x $10^{16}$ | -1.61 | 1860 | [42] |
| **$H_2 + CH_3O_2 \rightleftharpoons CH_2OH + CH_3O_2H$** | 1.50 x $10^{14}$ | 0.00 | 26030 | [38] |
| **$CH_3OH + CH_3O_2 \rightleftharpoons CH_2OH + CH_3O_2H$** | 1.81 x $10^{12}$ | 0.00 | 13710 | [49] |
| **$CH_3O_2H \rightleftharpoons CH_3O + OH$** | 6.31 x $10^{14}$ | 0.00 | 42300 | [42] |

## 3.2. ANALYSIS OF HYDROGEN DATASETS

There are only three published datasets for the IDT of hydrogen in $sCO_2$ from Shao et al. [9]. As seen in Table 6, Aramco 2.0 has the lowest average *E* value for the three hydrogen datasets but performs poorly at the H3 condition. This decline in the performance of all the mechanisms for the H3 dataset is possibly due to the lower pressure or the 0.25 equivalence ratio as opposed to 1.00 for the H1 and H2 datasets.

**Table 6.** Quantitative analysis of hydrogen datasets.

| Dataset | | | Average Absolute Error (*E*, (%)) | | | |
|---|---|---|---|---|---|---|
| **No.** | **Reference** | **Average Pressure /atm** | **Aramco 2.0** | **DTU** | **GRI 3.0** | **USC II** |
| **H1** | [9] | 109.6 | 18.7 | 35.4 | 71.7 | 44.5 |
| **H2** | [9] | 270.6 | 9.2 | 20.3 | 69.3 | 72.3 |
| **H3** | [9] | 38.4 | 85.5 | 111.5 | 123.1 | 75.5 |
| ***Average E (%)*** | | | ***37.8*** | ***55.7*** | ***88.0*** | ***64.1*** |
| ***No. Best Fit*** | | | ***2*** | ***0*** | ***0*** | ***1*** |

Two reactions which feature in the sensitivity analysis for H3 at 1274 K for Aramco 2.0 and USC II are R5 and R7 respectively as shown in Figure 3, showing the importance of $H_2O_2$ chemistry to hydrogen combustion. R5 was discussed previously due to its importance in methane combustion, however, R7 did not appear in any sensitivity analysis for methane combustion. The rate coefficient used in Aramco 2.0 for R7 [50] is significantly faster at the temperatures concerned compared to the value used in USC II [38], which is a possible explanation for the overprediction of the IDT in the H1 and H2 datasets by USC II.

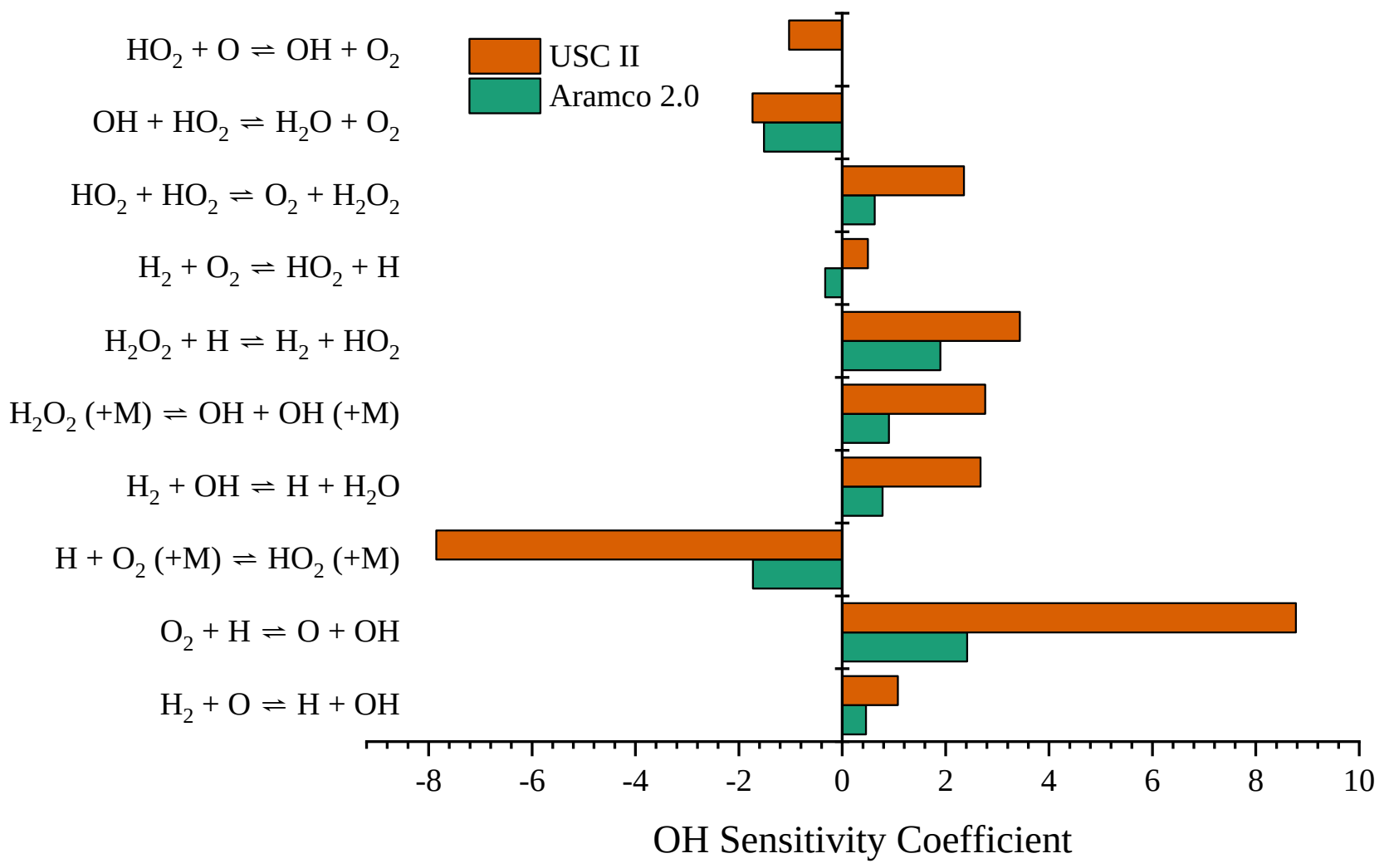

**Figure 3:** Plot of OH sensitivity analysis of H3 dataset of Aramco 2.0 and USC II at 1274 K.

As the experimental data can already be well modelled for H1 and H2 by Aramco 2.0, the challenge with the mechanism improvement is to greatly improve H3, without adversely affecting the aforementioned conditions. One of the most obvious discrepancies between the two mechanisms is the relative importance of the competing pathways of R8 and R9 in USC II. The older rate coefficient for R8 used by USC II from Masten et al. [51] is approximately 50 times larger than the rate coefficient reported by Hong et al. [52] used in Aramco 2.0. Conversely, the difference in the rate coefficient used in Aramco 2.0 [53] and USC II [54] only differ by 10%. The ratio between R8 and R9 is therefore important and influential on the combustion chemistry of hydrogen.

R7 $H_2O_2 + H \rightleftharpoons H_2 + HO_2$

R8 $O_2 + H \rightleftharpoons O + OH$

R9 $H + O_2 (+M) \rightleftharpoons HO_2 (+M)$

The sensitivity analysis for the H1 and H2 conditions in figure 3 shows a similar discrepancy between R8 and R9.

The mixed performance between Aramco 2.0 and USC II appears to be caused by the large difference in the rate coefficients used for the R7 and R8. The effect of altering the rate coefficient of R7 in the USC II mechanism to the faster, and more recent rate coefficient from Ellingson et al. [50] to create [USC II-Altered R7] is shown in Figure 4 for the H1 and H2 conditions. Furthermore, for the H3 condition, the *E* value falls from 85 and 75 for Aramco 2.0 and USC II respectively to 30 for [USC II-Altered R7].

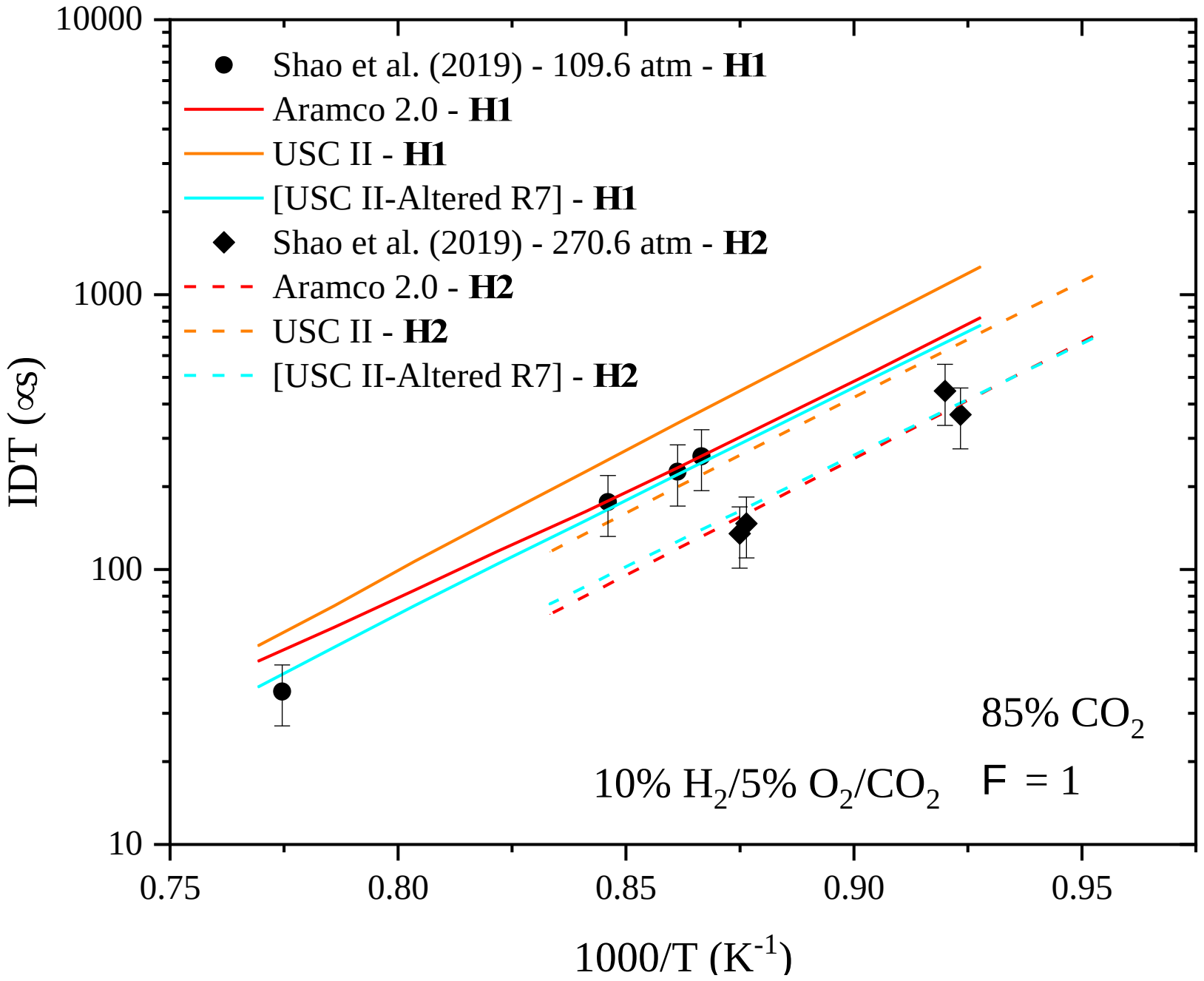

**Figure 4:** H1 and H2 dataset modelled by Aramco 2.0, USC II, and [USC II-Altered R7]: USC II with the updated R7 rate coefficient [50].

### 3.3. ANALYSIS OF SYNGAS DATASETS

Table 7 shows the average *E* values for the eight syngas datasets modelled. There is a significant increase in the overall average absolute error compared to methane and hydrogen with some *E* values for individual datasets being over 1000, with the mechanisms over predicting IDT in most datasets.

**Table 7.** Quantitative analysis of syngas data.

| Dataset | | | Average Absolute Error (*E*, (%)) | | | |
|---|---|---|---|---|---|---|
| **No.** | **Reference** | **Average Pressure /atm** | **Aramco 2.0** | **DTU** | **GRI 3.0** | **USC II** |
| **S1** | [18] | 1.2 | 82.4 | 104.8 | 43.2 | 17.4 |
| **S2** | [18] | 1.7 | 596.6 | 615.5 | 169.6 | 131.5 |
| **S3** | [18] | 2.3 | 451.3 | 498.3 | 64.2 | 45.7 |
| **S4** | [20] | 1.7 | 786.9 | 855.0 | 191.6 | 70.3 |
| **S5** | [20] | 1.7 | 1058.1 | 1067.4 | 233.9 | 81.5 |
| **S6** | [20] | 1.7 | 1194.3 | 1122.9 | 244.0 | 43.6 |
| **S7** | [20] | 1.7 | 578.2 | 552.3 | 147.8 | 38.6 |
| **S8** | [20] | 1.7 | 972.5 | 1030.9 | 430.7 | 121.7 |
| **S9** | [20] | 1.7 | 410.5 | 490.9 | 115.4 | 37.3 |
| **S10** | [21] | 41.5 | 185.8 | 238.6 | 127.1 | 179.8 |
| **S11** | [21] | 38.6 | 259.6 | 341.1 | 156.8 | 257.0 |
| **S12** | [21] | 38.5 | 191.6 | 277.5 | 49.8 | 128.9 |
| **S13** | [21] | 38.4 | 174.5 | 287.5 | 22.9 | 85.4 |
| **S14** | [17] | 78.9 | 43.3 | 58.1 | 37.9 | 40.2 |
| **S15** | [17] | 91.7 | 65.8 | 94.4 | 31.3 | 101.3 |
| **S16** | [17] | 89.6 | 117.7 | 180.8 | 39.8 | 115.5 |
| **S17** | [17] | 89.7 | 128.4 | 155.5 | 84.9 | 194.5 |
| **S18** | [19] | 101.0 | 31.2 | 57.8 | 52.4 | 45.4 |
| **S19** | [19] | 84.2 | 6.9 | 16.6 | 19.7 | 23.7 |
| **S20** | [19] | 208.0 | 10.8 | 5.1 | 11.2 | 4.5 |
| **Average *E*** | | | ***367.3*** | ***402.6*** | ***113.7*** | ***88.2*** |
| **No. Best Fit** | | | ***1*** | ***1*** | ***8*** | ***10*** |

The results from Table 7 clearly show that GRI 3.0 and USC II are the best performing mechanisms both in terms of the average *E* and the number of best fits. However, even the better performing USC II mechanism only fits 2 out of the 20 of the datasets within a 20% average absolute error, meaning that 90% of the datasets do not agree with the mechanism within a reasonable experimental error. Clearly, there is a need for significant improvement in a mechanism that can accurately replicate syngas ignition delay time data. Interestingly, the values with the largest absolute error are typically from the lowest pressure datasets S1-S9, with the highest-pressure dataset, S20, showing the best agreement between all four mechanisms and the experimental data. This is surprising given that these mechanisms are all validated for low-pressure combustion.

Figure 5 displays the sensitivity analysis for the S16 condition of Aramco 2.0 and GRI 3.0 at 1280 K. The discrepancy between the two mechanisms can be explained by the significant difference in the rate coefficient for R7 and R10.

R10 $CO + HO_2 \rightleftharpoons CO_2 + OH$

The rate coefficient of R10 used in GRI 3.0 [55] is significantly faster than that in Aramco 2.0 [56] leading to six orders of magnitude difference in the rate at 1200 K. Thus, explaining the much greater temperature sensitivity of R10 in GRI 3.0 compared to Aramco 2.0, DTU, and USC II which use the same rate coefficient.

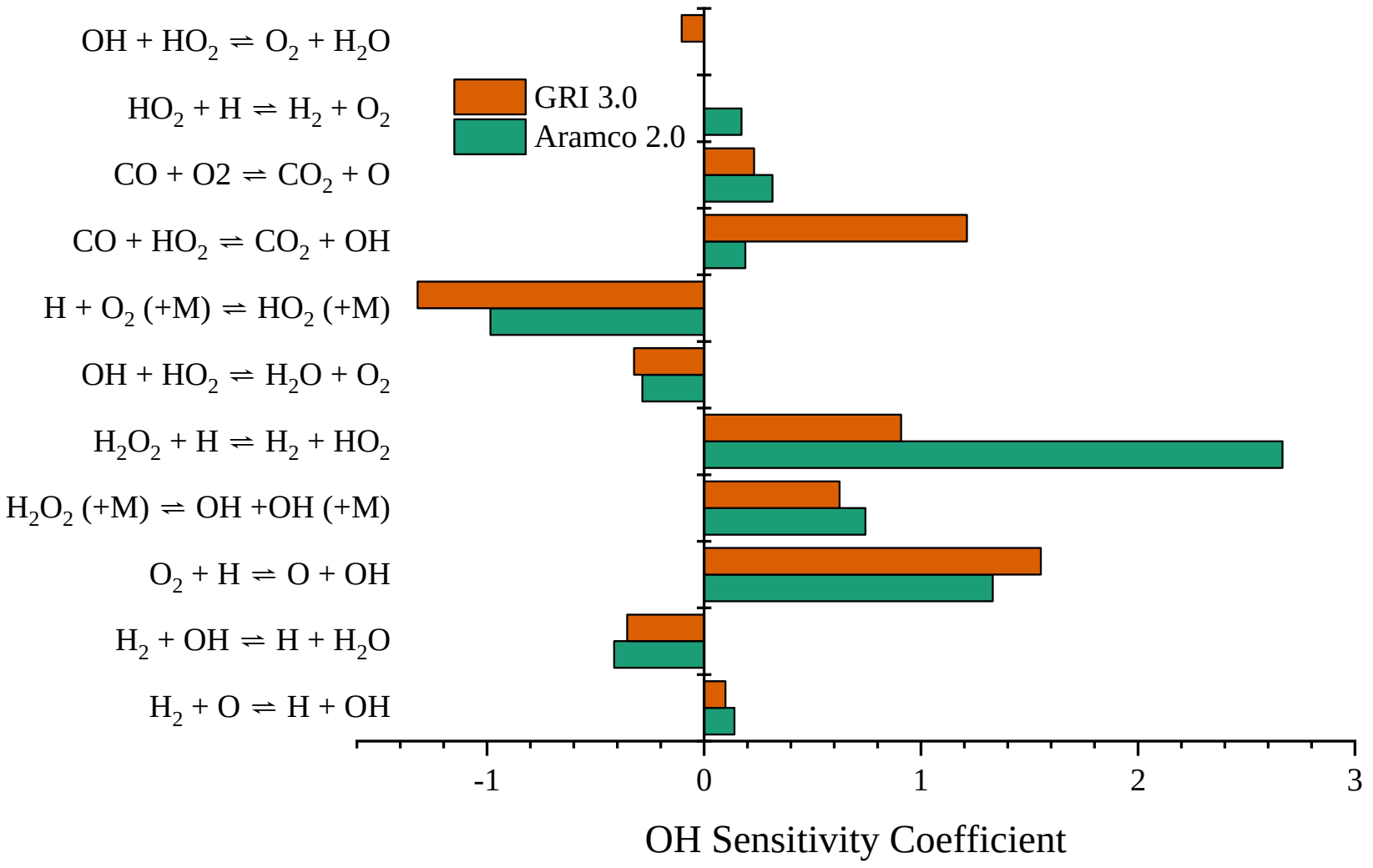

**Figure 5:** Plot OH sensitivity analysis of S16 dataset sensitivity analysis of Aramco 2.0 and GRI 3.0 1280 K.

Changing the rate coefficient for R10 for Aramco 2.0, DTU, and USC II to the faster rate coefficient led to a significant improvement for all three mechanisms across all the conditions as shown in Figure 6 for the S16 condition.

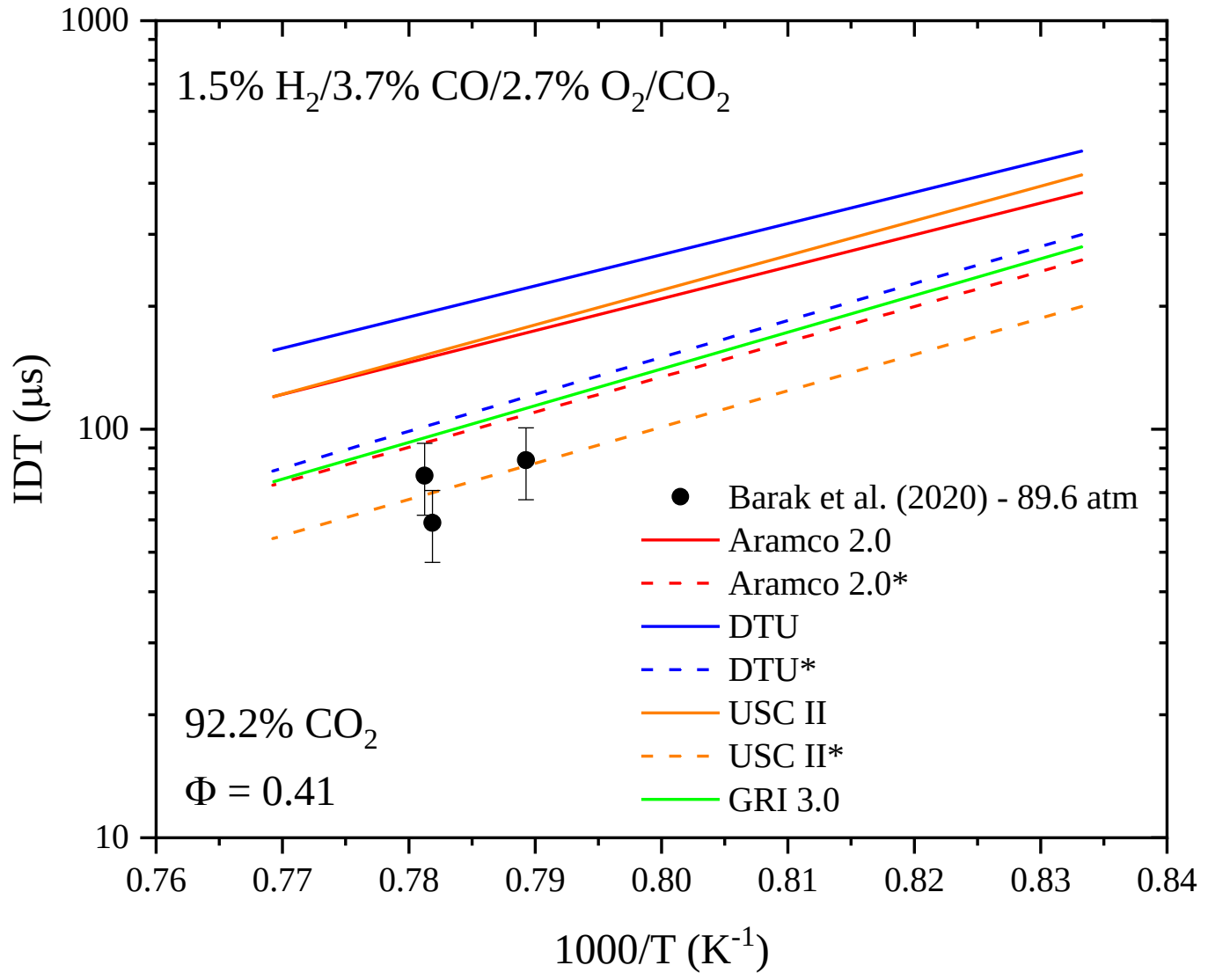

**Figure 6:** Mechanism comparison for the S16 dataset with the R10 rate coefficient from Baulch et al. [55] denoted by *.

One further observation is the difference between the syngas datasets at low and high-pressures. As shown in Figure 6, the datasets above 10 atm (S10-S20) exhibit a linear relationship with temperature on the log plot. Conversely, the low-pressure datasets have a curved IDT plot as shown in Figure 7 for the S2 dataset. It is in these conditions that Aramco 2.0 and DTU perform extremely poorly by overestimating the sharpness of the curvature, leading to a large discrepancy at low-temperature data points.

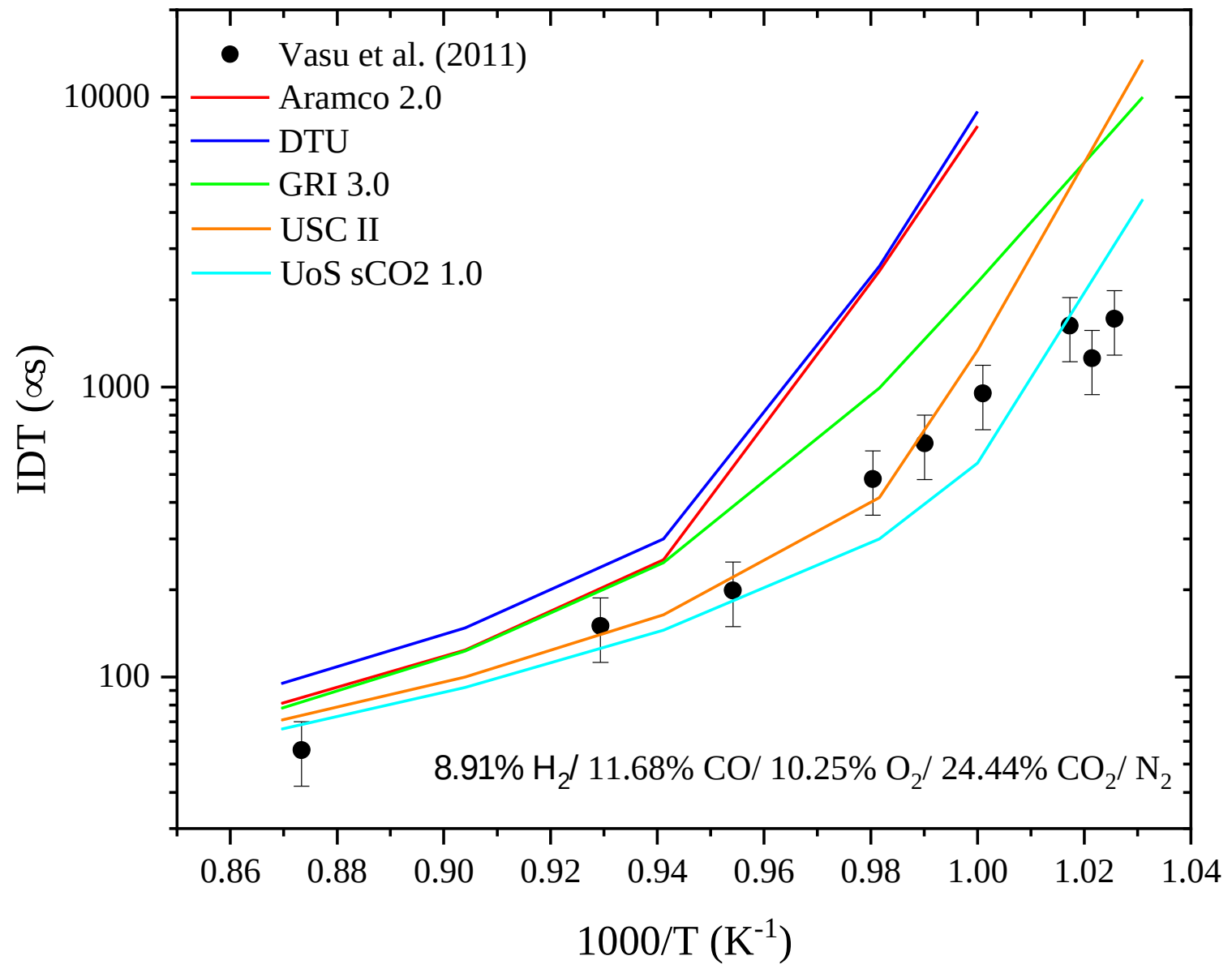

**Figure 7:** Mechanism comparison for the IDT of the S2 dataset.

A normalised sensitivity analysis of the S2 dataset at 1025 K and 1145 K is shown in Figure 8 shows the importance of the branching ratio between R8 and R9 is key to causing the sudden increase in IDT at 1025 K. As seen in Figure 8, the relative OH sensitivity coefficient at 1145 K for R9 is less than half that of R8, indicating the rate of the chain branching is far exceeding that of recombination to $HO_2$. At 1025 K, the relative OH sensitivity coefficients are similar, explaining the sudden and dramatic increase in IDT as the rate of radical reduction has slowed and the same is observed in the USC II sensitivity analysis which better models the experimental data. Therefore, it is again the importance of the chain branching ratio between R8 and R9 in determining the curvature in the IDT at lower temperatures.

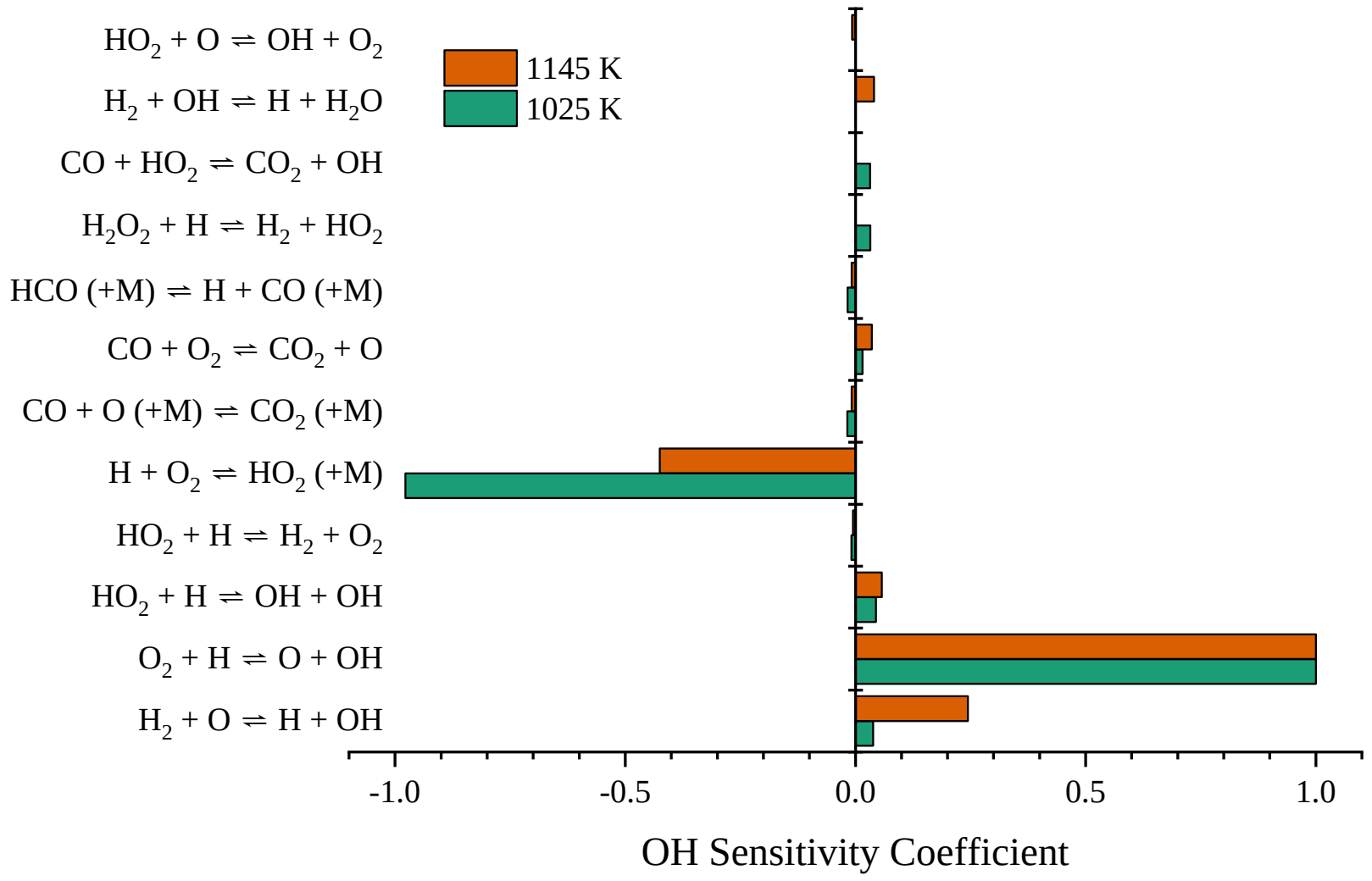

**Figure 8:** Plot of normalised OH sensitivity analysis of S2 data for Aramco 2.0 at 1025 and 1145 K.

## 4. CREATION OF THE UOS $sCO_2$ THE MECHANISM

The analysis of the three different fuels was used for the development of a new mechanism that better modelled all the datasets studied. USC II was chosen as the base mechanism, firstly due to it being the best performing mechanism for modelling syngas data which initially appeared to be the most difficult fuel to get a mechanism to model. Secondly, USC II showed a significant initial improvement for the high-pressure methane data through the addition of $CH_3O_2$ chemistry shown in Figure 2, whilst maintaining its superior performance at lower pressure, this meant it quickly became a much more competitive mechanism for the methane datasets. The modelling loop performed for the creation of the UoS $sCO_2$ mechanism is shown in Figure 9. Conditions where the UoS $sCO_2$ mechanism was identified using the *E* (%) value and subsequently investigated using sensitivity analysis at points of discrepancy between experimental and simulated datapoints. The sensitivity analyses generated were used to identify reactions important at these conditions, and their rate coefficients were investigated, to see what alterations could be made within reason, or switch to a rate coefficient used in another mechanism. It was also important at this stage to investigate other datasets in good agreement with UoS sCO2, to check that any changes to these rate coefficients would not have adverse effects on these datasets. The new mechanism was then checked across all the datasets, ensuring that the average *E* (%) value across all datasets dropped. In order, this was repeated for hydrogen, syngas, and methane, always checking that any changes did not have any adverse effect on the former fuels. The reason this order was selected for the fuels is two-fold. Firstly, to increase the complexity of the fuels as the mechanism progressed, meaning that later changes to methane chemistry, had no, or very little effect, on the former fuels. Secondly, it was performed in an increasing number of datasets, meaning that the *E* (%) value had to be re-evaluated for fewer fuels when comparing the first few datasets.

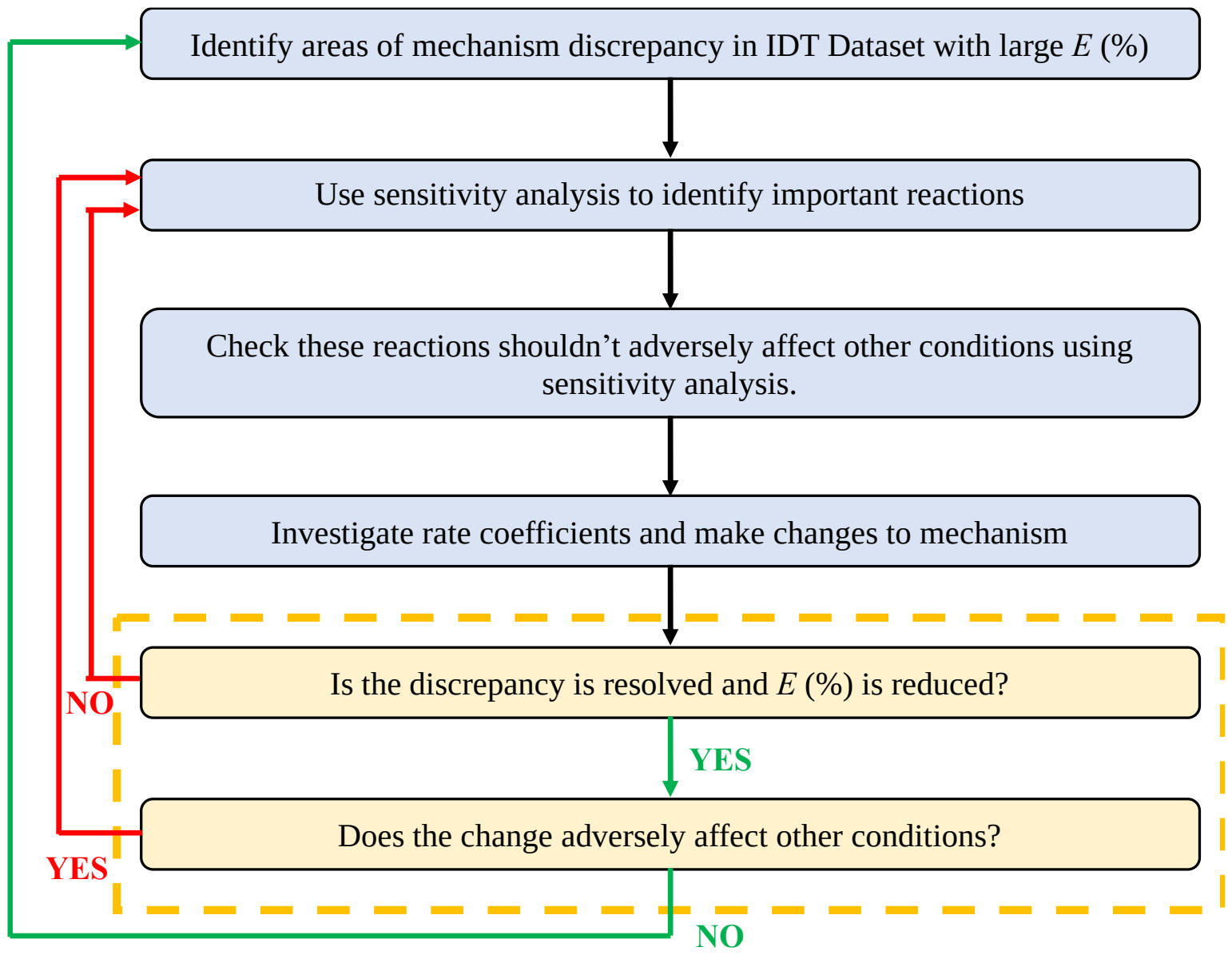

**Figure 9:** Modelling loop performed for the creation of the UoS $sCO_2$ mechanism.

Table 5 shows the $CH_3O_2$ reactions and rate coefficients taken from the Aramco 2.0 mechanism which were added to USC II. All other changes to the rate coefficients are recorded in Table 8 along with the reference to the rate coefficients and the magnitude of the change.

**Table 8.** UoS $sCO_2$ mechanism alterations.

| Reaction | Mechanism | A ($cm^3$ mol s) | n | Ea (cal/mol) | Reference |
|---|---|---|---|---|---|
| $H_2O_2 + H \rightleftharpoons HO_2 + H_2$ | USC II | $6.1 \times 10^{6}$ | 2 | 5200 | [38]*0.5 |
| | New | $1.85 \times 10^{10}$ | 1 | 6000 | [50]*1.2 |
| $CO + O_2 \rightleftharpoons CO_2 + O$ | USC II | $1.1 \times 10^{12}$ | 0 | 47700 | [38] |
| | New | $2.5 \times 10^{12}$ | 0 | 47700 | [38]*2.27 |
| $CO + HO_2 \rightleftharpoons CO_2 + OH$ | USC II | $1.6 \times 10^{5}$ | 2.18 | 17940 | [56] |
| | New | $4 \times 10^{5}$ | 2.18 | 17942 | [56]*2.5 |
| $CH_3 + O_2 \rightleftharpoons O + CH_3O$ | USC II | $3.1 \times 10^{13}$ | 0 | 28800 | [57] |
| | New | $1.0 \times 10^{13}$ | 0 | 28320 | [58]*1.3 |
| $CH_3 + O_2 \rightleftharpoons OH + CH_2O$ | USC II | $3.6 \times 10^{10}$ | 0 | 8940 | [57] |
| | New | $1.7 \times 10^{11}$ | 0 | 9842 | [58]*0.9 |
| $CH_3 + HO_2 \rightleftharpoons CH_4 + O_2$ | USC II | $1.0 \times 10^{12}$ | 0 | 0 | [46] |
| | New | $1.2 \times 10^{5}$ | 2.23 | -3022 | [47] |
| $CH_3 + HO_2 \rightleftharpoons CH_3O + OH$ | USC II | $1.3 \times 10^{13}$ | 0 | 0 | [38] |
| | New | $1.10 \times 10^{12}$ | 0.269 | -687.5 | [47] |
| $CH_3 + H_2O_2 \rightleftharpoons CH_4 + HO_2$ | USC II | $2.5 \times 10^{4}$ | 2.47 | 5180 | [59] |
| | New (R) | $4.7 \times 10^{4}$ | 2.5 | 21000 | [60] |
| $CH_4 + H \rightleftharpoons CH_3 + H_2$ | USC II | $6.6 \times 10^{8}$ | 1.62 | 10840 | [61] |
| | New | $6.1 \times 10^{5}$ | 2.5 | 9587 | [60] |
| $H + O_2 \rightleftharpoons O + OH$ | USC II | $2.644 \times 10^{16}$ | -0.6707 | 17041 | [62] |
| | New | $3.0 \times 10^{16}$ | -0.6707 | 17041 | [62]*1.13 |
| OH + H2 ⇌ H + H2O | USC II | $1.734 \times 10^{8}$ | 1.51 | 3430 | [63] |
| | New | $4.0 \times 10^{8}$ | 1.51 | 3430 | [63]*2.31 |
| $H + O_2$ (+M) ⇌ $HO_2$ (+M) | USC II | $5.116 \times 10^{12}$ | 0.44 | 0 | [54]*1.1 |
| | New | $3.00 \times 10^{12}$ | 0.44 | 0 | [54]*0.64 |
| $H_2 + O_2 \rightleftharpoons HO_2 + H$ | USC II | $5.916 \times 10^{5}$ | 2.433 | 53502 | [64]*0.8 |
| | New | $4.5 \times 10^{5}$ | 2.433 | 53502 | [64]*0.61 |
| OH + OH (+M) ⇌ $H_2O_2$ (+M) | USC II | $1.11 \times 10^{14}$ | -0.37 | 0 | [65] |
| | New | $9.5 \times 10^{13}$ | -0.37 | 0 | [65]*0.86 |
| $HO_2 + H \rightleftharpoons OH + OH$ | USC II | $7.485 \times 10^{13}$ | 0 | 295 | [66-68] |
| | New | $4.5 \times 10^{13}$ | 0 | 295 | [66-68]*0.6 |
| $HO_2 + O \rightleftharpoons OH + O_2$ | USC II | $4 \times 10^{13}$ | 0 | 0 | [67]*2 |
| | New | $2 \times 10^{13}$ | 0 | 0 | [67] |
| $H_2O_2 + OH \rightleftharpoons HO_2 + H_2O$ (DUP) | USC II | $2 \times 10^{12}$ | 0 | 427 | [69] |
| | New | $1 \times 10^{12}$ | 0 | 427 | [69]*0.5 |
| $H_2O_2 + OH \rightleftharpoons HO_2 + H_2O$ (DUP) | USC II | $2.67 \times 10^{41}$ | -7 | 37600 | [69] |
| | New | $1.5 \times 10^{41}$ | -7 | 37600 | [69]*0.56 |
| $CO + OH = CO_2 + H$ (DUP) | USC II | $7.04 \times 10^{4}$ | 2.053 | -355.67 | [70] |
| | New | $9 \times 10^{4}$ | 2.053 | -355.67 | [70]*1.28 |
| $CO + OH \rightleftharpoons CO_2 + H$ (DUP) | USC II | $5.757 \times 10^{12}$ | -0.664 | 331.83 | [70] |
| | New | $7.5 \times 10^{12}$ | -0.664 | 331.83 | [70]*1.30 |
| $OH + HO_2 \rightleftharpoons H_2O + O_2$ (DUP) | USC II | $1.41 \times 10^{18}$ | -1.76 | 60 | [32] |
| | New | $2.8 \times 10^{18}$ | -1.76 | 60 | [32]*2 |
| $OH + HO_2 \rightleftharpoons H_2O + O_2$ (DUP) | USC II | $1.12 \times 10^{85}$ | -22.3 | 26900 | [32] |
| | New | $2.24 \times 10^{85}$ | -22.3 | 26900 | [32]*2 |
| $OH + HO_2 \rightleftharpoons H_2O + O_2$ (DUP) | USC II | $5.37 \times 10^{70}$ | -16.72 | 32900 | [32] |
| | New | $1.2 \times 10^{71}$ | -16.72 | 32900 | [32]*2.23 |
| $OH + HO_2 \rightleftharpoons H_2O + O_2$ (DUP) | USC II | $2.51 \times 10^{12}$ | 2 | 40000 | [32] |
| | New | $5 \times 10^{12}$ | 2 | 40000 | [32]*2 |
| $OH + HO_2 \rightleftharpoons H_2O + O_2$ (DUP) | USC II | 1.00 x $10^{136}$ | -40 | 34800 | [32] |
| | New | 2 x$10^{136}$ | -40 | 34800 | [32]*2 |
| $HCO (+M) \rightleftharpoons CO + H (+M)$ | USC II | 1.87 x $10^{17}$ | -1 | 17000 | [71]*2 |
| | New | 4 x $10^{17}$ | -1 | 17000 | [71]*4.28 |
| $CH_3 + OH \rightleftharpoons CH_2^* + H2O$ | USC II | 2.501 x $10^{13}$ | 0 | 0 | [72-77] |
| | New | 1.75 x $10^{13}$ | 0 | 0 | [72-77]*0.7 |
| $CH_2^* + CO_2 \rightleftharpoons CH_2O + CO$ | USC II | 1.4 x $10^{13}$ | 0 | 0 | [78] |
| | New | 7.0 x $10^{12}$ | 0 | 0 | [78]*0.5 |
| $CH_3 + CH_3 \rightleftharpoons C_2H_5 + H$ | USC II | 4.99 x $10^{12}$ | 0.1 | 10600 | [79] |
| | New | 7.5 x $10^{12}$ | 0.1 | 10600 | [79]*1.5 |
| $CH_2O + O_2 \rightleftharpoons HCO + HO_2$ | USC II | 1.00 x $10^{14}$ | 0 | 40000 | [80] |
| | New | 1.50 x $10^{14}$ | 0 | 40000 | [80]*1.5 |

Table 9 shows the performance of the newly developed UoS $sCO_2$ mechanism against the four existing mechanisms studied for hydrogen, methane, and syngas data.

**Table 9.** Comparison of the developed UoS $sCO_2$ mechanism to the existing four mechanisms studied.

| **Fuel** | | **Mechanism** | | | | |
|---|---|---|---|---|---|---|
| | | **Aramco 2.0** | **DTU** | **GRI 3.0** | **USC II** | **UoS $sCO_2$** |
| **Hydrogen** | *Average E* (%) | 37.8 | 55.7 | 88.0 | 64.1 | 11.7 |
| | No. Best Fit | 1 | 0 | 0 | 0 | 2 |
| **Methane** | *Average E* (%) | 28.9 | 23.9 | 32.7 | 39.2 | 17.5 |
| | No. Best Fit | 2 | 4 | 5 | 5 | 13 |
| **Syngas** | *Average E* (%) | 367.3 | 402.6 | 113.7 | 88.2 | 36.2 |
| | No. Best Fit | 1 | 0 | 2 | 2 | 15 |
| **Average *E*** | | ***144.7*** | ***160.7*** | ***78.1*** | ***63.8*** | ***21.8*** |
| **Total No. Best Fit** | | ***4*** | ***4*** | ***7*** | ***7*** | ***30*** |

Firstly, the UoS $sCO_2$ mechanism shows a significant reduction in the average *E* value of 11.7%, given most shock tube IDTs have an error of approximately 20%, this can be considered to be a good mechanism performance. UoS $sCO_2$ is also the best fit to two of the three datasets, fitting all the datasets with an average *E* of less than 20% in all three datasets. The problem is the lack of experimental data for hydrogen IDTs in $CO_2$. More datasets are required to fill in the gaps between the three average pressures studied and look at different equivalence ratios and $CO_2$ concentrations.

For the methane IDT data, UoS $sCO_2$ has reduced the average *E* value to 17.5% from 23.9%, the next best performing mechanism DTU. Furthermore, UoS $sCO_2$ is the best fitting mechanism to thirteen of the twenty-nine datasets, fitting 22 of the datasets within an average *E* of 20%. Unlike the performance with hydrogen fuel, there is a vast number of datasets covering a wide range of pressures and conditions. This provides significant confidence that the UoS $sCO_2$ mechanism can model the IDT of methane in oxyfuel conditions within an *E* (%) value of 17.5 ± 12.1%. The error was calculated using the standard deviation.

The existing mechanisms used in the study struggled to model the syngas IDT data, with Aramco 2.0 and DTU both having average *E* (%) values above 350%. As previously discussed, this was predominantly due to an overestimation in the ignition delay time and the simulation lacking the correct curvature of the low-pressure datasets. The UoS $sCO_2$ average *E (%)* value of 36.2 ± 24.5% is a significant improvement on the existing mechanisms in addition to being the best fit to 75% of the twenty datasets studied. However, the average *E* for UoS $sCO_2$ for syngas IDT data is still over double that of methane and hydrogen, and therefore there is still some room for improvement in the worse performing datasets. The challenge is doing so without adversely affecting the better fitting datasets.

It should be noted that the smallest *E* value does not necessarily indicate the best fit due to the large error of 18-25% in the IDT, meaning multiple mechanisms may be within the error. However, across a large number of datasets, the average *E* value is a good indication of the best performing mechanism. Table 9 shows the UoS $sCO_2$ mechanism produces the best average *E* for all of the fuels studied and is the best fit to over half of the datasets. It is important to note that the rate coefficients selected are not a reflection of the quality of the rate coefficient as they were not measured in a $CO_2$ bath gas, the selection was based on creating the best fit to the available experimental data available.

## 5. CONCLUSION

The UoS $sCO_2$ has been proven to better model the IDT of methane, hydrogen, and syngas oxyfuel combustion at various pressures ranging from sub-atmospheric to over 250 atm, and various $CO_2$ dilutions and equivalence ratios. Through studying 52 datasets taken from 12 different publications, the UoS $sCO_2$ has been created through the reasonable modifications to reaction rate coefficients from the USC II base mechanism as identified through sensitivity analysis. The development of a mechanism specific to oxy-fuel combustion is essential to accurately model combustion in direct-fired $sCO_2$ power cycles such as the Allam-Fetvedt and STEP.

The importance of the $CH_3O_2$ chemistry in high-pressure methane combustion was discussed for the first time with relevance to the Allam cycle. It was noted during the methane analysis that the GRI 3.0 and USC II mechanisms performed poorly at high-pressure conditions due to the absence of the chemistry of $CH_3O_2$. The addition of the $CH_3O_2$ chemistry from Aramco 2.0 to USC II saw a significant improvement in the three highest pressure conditions. However, most of these reactions have been studied concerning atmospheric and low-temperature combustion so may be unreliable when extrapolated to the conditions of the Allam-Fetvedt cycle.

The rate coefficients used in the creation of the UoS $sCO_2$ mechanism were based on which best modelled the experimental data. Alterations to the rate coefficients were made within a reasonable assumption of error based on the literature source. More high-pressure combustion data for methane, hydrogen and syngas is required to confirm the changes made to rate coefficients selected and validate the mechanism. This study furthers the current understanding of $sCO_2$ combustion of methane, hydrogen, and syngas by assessing the performance of four combustion mechanisms well-validated at lower pressures and not in $CO_2$. Quantitative analysis of the newly developed UoS $sCO_2$ mechanism adapted from USC II has proven its superior ability to model IDT against existing chemical kinetic mechanisms. Whilst the UoS $sCO_2$ mechanism has been proven to perform best for the IDTs studied, more experimental data such as laminar flame speed is required for further validation that the mechanism better models all aspects of combustion. Overall, the present study allows the IDT of three different fuels to be more accurately modelled at conditions of the direct-fired $sCO_2$ combustion than by four existing, well-established chemical kinetic mechanisms across 52 experimental datasets. This mechanism can henceforth be used to better model the high-pressure combustion of both methane and syngas at the conditions of the Allam-Fetvedt cycle, the information gained from which can be used to improve the efficiency of the combustion chamber.

## ACKNOWLEDGEMENTS

This work has been supported by the EPSRC Centre for Doctoral Training in Resilient Decarbonised Fuel Energy Systems (Grant number: EP/S022996/1) and the International Flame Research Federation (IFRF).